\documentclass[[journal]{IEEEtran}
\usepackage{cite}
\usepackage{amsmath,amssymb,amsfonts}
\usepackage{algorithmic}
\usepackage{graphicx}
\usepackage{textcomp}
\usepackage{float}
\usepackage{subfloat}
\usepackage{placeins}
\usepackage{caption}
\usepackage{subcaption}

\usepackage{color,soul}

\begin{document}
\title{PER Measurement of BLE in RF Interference and Harsh Electromagnetic Environment}
\author{Mir~Lodro, Gabriele~Gradoni , Ana~Vukovic, David~Thomas and Steve~Greedy  \vspace{-0.75cm}
\thanks{Mir Lodro, Steve Greedy, Ana Vukovic, David Thomas and Gabriele Gradoni are with George Green Institute for Electromagnetic Research-GGIEMR, the University of Nottingham, UK. Gabriele Gradoni is also with British Telecommunications and the University of Cambridge, UK.}}
\maketitle
\begin{abstract}
Bluetooth Low Energy (BLE) is a short-range data transmission technology that is used for multimedia file sharing, home automation, and internet-of-things application. In this work, we perform packet error rate (PER) measurement and RF testing of BLE receiver in the harsh electromagnetic environment and in presence of RF interference. We check the PER performance in the line-of-sight (LOS) and non-line-of-sight (NLOS) scenario in absence of any interfering signal and in presence of wideband WLAN interference. The BLE PER measurements are conducted in a large reverberation chamber which is a rich scattering environment. Software-defined-radio has been used to create BLE communication link for PER measurement in LOS and NLOS configuration. The BLE PER is measured both in the presence and in absence of WLAN interference. Our measurement results show a higher PER for uncoded BLE PHY modes in NLOS channel condition and in presence of wideband interference. Whereas coded BLE PHY modes i.e. LE500K and LE125K are robust to interference with lower PER measurements.  
\end{abstract}

\vspace{-0.12cm}

\vspace{-0.35cm}
\section{Introduction}
Bluetooth low energy (BLE) has become critically important as it has been demonstrated by authors in \cite{sattler2020risk}\cite{cunche2020using}\cite{leith2020coronavirus} where it is shown that BLE can potentially contain and delay the outbreak of infectious diseases such as the ongoing severe acute respiratory syndrome coronavirus 2 (SARS-CoV-2) pandemic. BLE is a wireless personal area network (WPAN) technology created by Bluetooth special interest group (SIG) as BLE 4.0. The latest BLE 5.0 can attain a data rate of 2 Mbps with two additional coded PHY modes supporting a data rate of 500 kbps and 125 kbps respectively. In contrast to classical Bluetooth, BLE is targeted for short-range power-constraint internet of things (IoT) applications such as medical internet of things, smart homes, indoor positioning, and localization \cite{montoliu2020indoor}\cite{spachos2020ble}. BLE operates in two modes connected communication and non-connected communication. Non-connection mode is used for device discovery to make a connection for data transfer. It's also known as the neighbor discovery process (NDP). Once the link is established using the NDP process, the two devices can perform data transfer using connected communication mode.
BLE is a low-power short-range communication system. Apart from multimedia file transmission between mobile phones its usage in IoT applications and home automation is on the constant rise. BLE backscatter communication is presented in \cite{ensworth2017ble}\cite{ensworth2017full}. Work in \cite{yu2020plant} shows smart city soil health monitoring for healthy plants. Authors in \cite{la2018dense} have addressed dense deployment of BLE for the body are networks. BLE neighbor discovery issue is addressed in \cite{luo2019ble}. The rapid influx of internet of things devices per square kilometer has increased the risk of interference in an already spectrally contested EM environment. IoT devices operate using WiFi, Zigbee, BLE, and narrowband-IoT (NB-IoT) PHY layers. Almost all IoT devices use the most sought-after 2.4 GHz ISM band. This paper is organized into five sections.
Section I is the introduction, Section II is the background, Section III is about measurement setup. Experimental results are presented in Section IV. Conclusion and future aspect of the work has been given in conclusion in Section V.
\vspace{-0.30cm}
\section{Background}
BLE uses GMSK for LE1M and LE2M PHY modes that support uncoded 1 Mbps and 2 Mbps data rates respectively. BLE also uses coded LE500K and LE125K PHY layer modes that support data rates of 500 kbps and 125 kbps respectively. BLE has four maximum transmit power classes. Class 1, class 1.5, class 2 and class 3 allow maximum transmit power of 20 dBm (100 mW), 10 dBm (10 mW), 4 dBm (4 mW) and 0 dBm (1 mW) respectively. BLE uses GMSK modulation in comparison to basic rate/enhanced data rate (BR/EDR) Bluetooth which used GFSK, $\pi/4$-DQPSK, and 8-DPSK modulation techniques. Additionally, BLE uses broadcast, mesh, and point-to-point network topologies.
\vspace{-0.30cm}
\subsection{BLE Frequency Plan}
\begin{figure}
    \centering
    \includegraphics[width=\columnwidth]{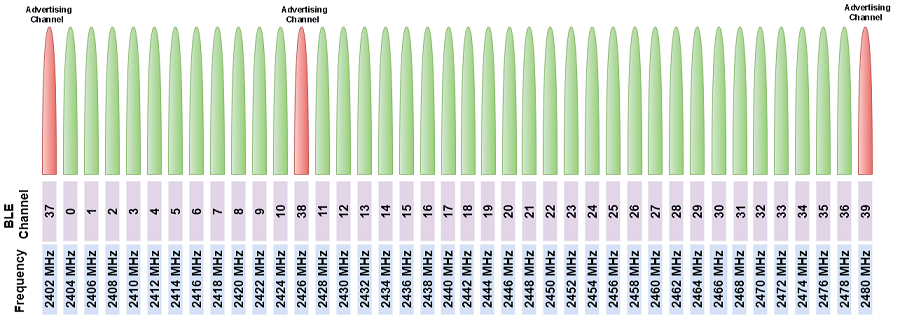}
    \caption{BLE frequency plan showing advertising channels and data channels.}
    \label{fig:ble_plan}
\end{figure}

 The BLE system uses forty RF channels separated by 2MHz, unlike Zigbee which uses sixteen RF channels separated by 5 MHz. BLE uses 2.4 GHz ISM band at 2400-2483.5 MHz. Figure \ref{fig:ble_plan} shows BLE RF channels where each channel is allocated a unique channel index. The BLE frequency plan shows advertising and data channels. BLE has two main channels advertising channels and data channels. Advertising channels are unidirectional or broadcast channels for device discovery and beacon signals. Out of forty BLE, RF channels three channels (channel number 37, 38, and 39) are advertising channels and thirty-seven channels (channel 0-36) are data channels. Advertising channels are spread across BLE 2.4 GHz ISM band in order to avoid interference. The purpose of advertising channels is to transmit advertising packets, scan packets, and connection packets.
\vspace{-0.30cm}
\subsection{BLE Channel Selection Algorithms}
BLE uses channel hopping to avoid interference and yield maximum throughput. 2.4 GHz ISM band is used by several wireless standards such as WLAN, therefore
BLE is susceptible to interference. In presence of interference, many retransmission may be requested for successful data delivery. BLE advertising device transmits advertising packets on three advertising channels in a cyclic manner.
The scanning device listens on three advertising channels in a cyclic manner. There are two BLE channel selection algorithms whose purpose is to avoid channels that are susceptible to interference. Good and bad data channel map is selected based on key performance indicators (KPIs), for example, SNR and PER. The channel map is exchanged between the master and the slave. Algorithm1 is governed by hope count, current channel, and a list of used good channels. Algorithm2 uses access address and a list of used good channels as main parameters.
\vspace{-0.35cm}
\subsection{Packet Format}
\begin{figure}
    \centering
    \subfloat[Packet format BLE uncoded]{\label{fig:ble_uncoded}\includegraphics[width=\columnwidth]{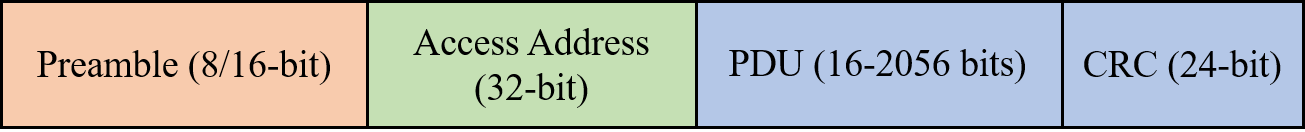}}\\
    \subfloat[Packet format BLE coded]{\label{fig:ble_coded}\includegraphics[width=\columnwidth]{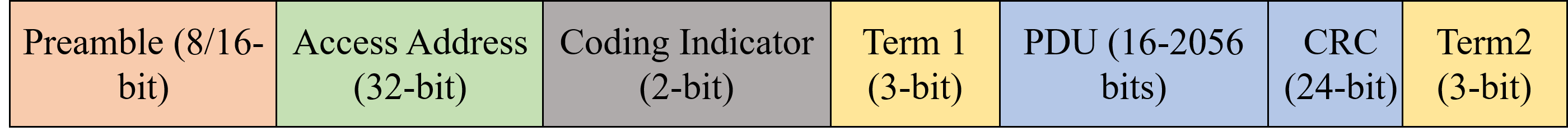}}
    \caption{BLE packet format structure for uncoded and coded PHY modes.}
    \label{fig:pkt_struct}
\end{figure}

BLE has two packet formats for both uncoded and coded PHY layers. The packet format for uncoded LE1M and LE2M PHY modes of BLE is shown in Fig.\ref{fig:ble_uncoded}. The preamble is a fixed sequence of alternating 1 and 0 bit, and it is part of all link-layer BLE packets. The preamble is used in the receiver to perform functions like automatic gain control (AGC) function, frequency synchronization, and timing recovery. The preamble length for LE1M PHY and LE2M PHY modes is 8 bits and 16 bits respectively. It consists of a 32-bit access address, 16-2056 bit PDU, and 24-bit CRC. The packet format for coded PHY LE125K and LE500K is shown in Fig.\ref{fig:ble_coded}. It has an 80-bit preamble, 32-bit access address, 16-2056 PDU bits, and 24-bit CRC. It has also three more fields 2-bit coding indicator that indicates coding level hence coded BLE PHY layer and 3-bit term1 and term 2 fields.

\begin{table}[]
    \centering
    \caption{BLE SPECIFICATIONS}
    \begin{tabular}{|c|c|}
    \hline
         Parameters&Value\\
      \hline
         Frequency Range&2.4-2.4835 GHz  \\
         Modulation& GMSK\\
         PHY layer modes& LE1M,LE2M,LE500K,LE125K\\
         Number of Channels& 40\\
         Frame Length& 16832\\
         Total number of frames&10,000\\
         \hline
    \end{tabular}
   
    \label{tab:ble_specifications}
\end{table}
\vspace{-0.35cm}
\subsection{BLE Waveform Generation}
\begin{figure}
    \centering
    \subfloat[]{\label{fig:uncoded_wave}\includegraphics[width=\columnwidth]{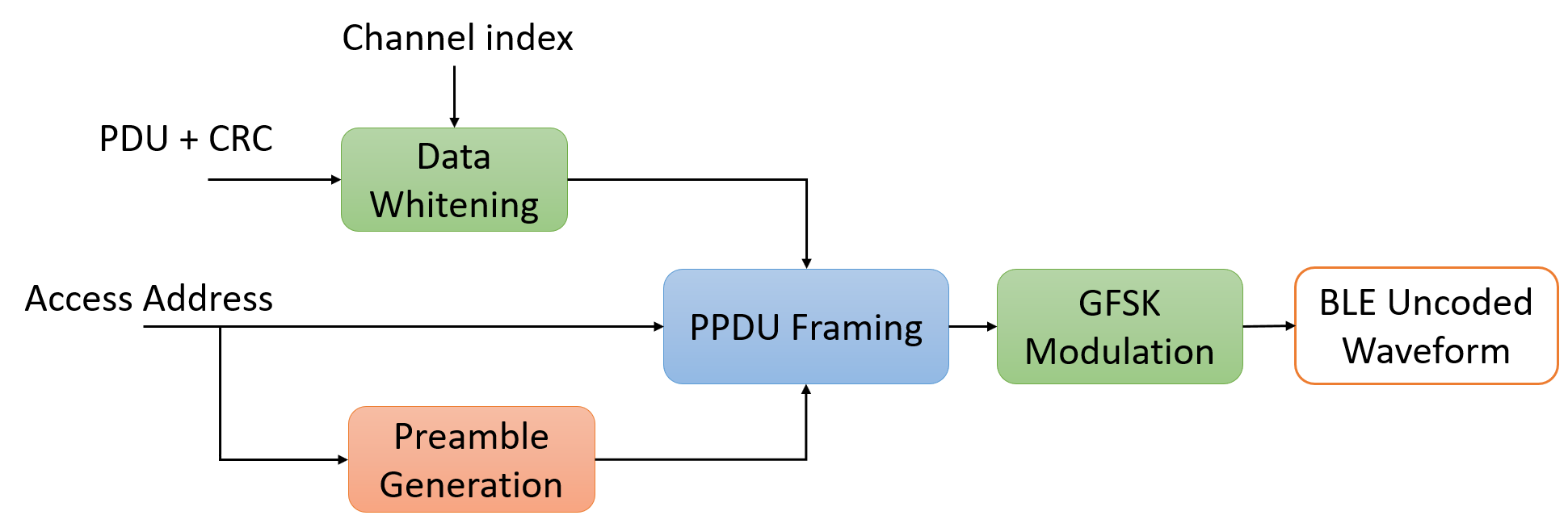}}\\
    \subfloat[]{\label{fig:coded_wave}\includegraphics[width=\columnwidth]{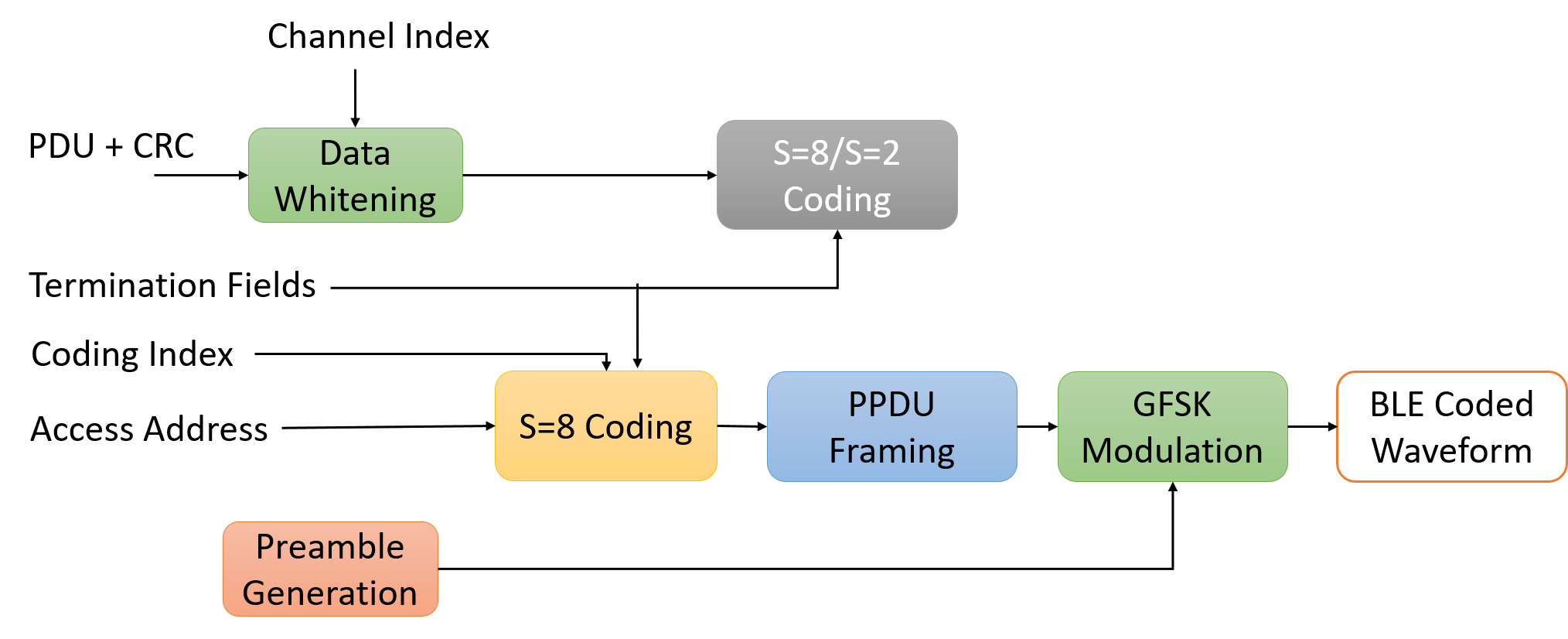}}
    \caption{BLE uncoded and coded waveform generation.}
    \label{fig:ble_procedure}
\end{figure}

BLE link layer PDUs both advertising PDUs and data channel PDUs can be generated and can be converted into BLE waveforms as shown in Fig.\ref{fig:ble_procedure}. PDUs are converted to standard-compliant baseband waveforms which are transmitted over the air using a software-defined-radio platform. Fig. \ref{fig:uncoded_wave} and Fig.\ref{fig:coded_wave} shows procedure for BLE waveform generation of uncoded and coded PHY modes respectively.
\vspace{-0.35cm}
\subsection{BLE Receiver}
BLE receiver performs RF impairment compensation such as RF nonlinearity effect,
carrier frequency offset, carrier phase offset, and symbol timing synchronization. Standard-compliant BLE receivers consist of AGC stage, DC compensation stage, matched filtering, carrier frequency offset (CFO) estimation, and timing synchronization stage. After the initial receiver operations demodulation of GMSK symbols is accomplished and a CRC check is performed. The AGC component in the digital receiver is vital for the correct recovery of information and for end-to-end PER performance measurement. The AGC is the first stage where a weak signal is received and stabilized for further downstream processing. The AGC can be configured in slow-attack mode or fast-attack mode to track the received signal. Fast attack AGC is used when the environment is non-stationary. Previously we have recorded our experimental experience with AGC in the metal enclosure \cite{lodro2020near}. The AGC helps for coarse and fine frequency correction in the baseband receiver. After the AGC stage, DC removal in the received waveform is performed using a notch filter. The existence of the DC component is a common problem in radio receivers and it is particularly common in direct-conversion receivers such as software-defined-radios. The next two stages are frequency offset correction and matched filtering. Frequency offset estimation and correction must be performed before received samples are processed by a matched filter. The matched filter is implemented half at the transmitter and a half at the receiver and it increases the SNR i.e. it limits the amount of noise that is passed down to subsequent stages\cite{litwin2001matched} in the receiver. Coarse frequency compensation uses an FFT-based or correlation-based algorithm to measure frequency offset. Fine frequency and demodulation of GMSK packets are performed. After these critical design stages, data decoding and de-mapping are performed. Long series of zeros and ones such as 000000 or 1111111 may occur in the data bitstream. Data whitening is used to avoid such a long sequence of ones and zeros. Data de-whitening and CRC are performed to measure the BLE PER. For the packet error measurement, access address is checked first and then CRC is checked, If the access address or CRC is incorrect then the packet shall be rejected. The received BLE packets is considered only valid when both access address and CRC check is correct.
\vspace{-0.35cm}
\section{Measurement Setup}
We generated a baseband BLE waveform using MATLAB based BLE transmitter. The generated BLE waveforms were transmitted using PlutoSDR connected to the host PC. The transmitted BLE signals were received by another PlutoSDR connected with the same host PC running the BLE receiver in another MATLAB session. PlutoSDR is a full-duplex transceiver. PlutoSDR uses Xilinx Zynq Z-7010 FPGA and a highly integrated RF agile transceiver chip AD9363 from Analog Devices. It has RF coverage from 325 MHz to 3.8 GHz which is software upgradeable from 70 MHz to 6 GHz. It can provide instantaneous bandwidth of 20 MHz and it offers 12-bit ADC/DAC flexible rate. It has support from MATLAB, Simulink, and GNU Radio and it has C++ and Python API. We performed BLE PER measurement in NDP mode in LOS and NLOS scenario in reverberation chamber which is a highly reflecting environment. In RC the scattering is very rich, therefore, a higher number of multipath components can be expected similar to a factory environment where dense metal bars, enclosures, ducts, and other instrumentation are used. We recorded higher PER in the reverberation chamber for Tx gain of 0 dB and Rx gain of 25 dB. We repeated the same with Tx and Rx gain values in an indoor laboratory environment where PER of $1e^{-2}$ was measured with repeatable measurements. The PER measurements within RC were around $0.5$ with repeatable measurements, We also ruled out any possibility of cables connecting PlutoSDRs with host PC were faulty.
\begin{figure}
    \centering
    \includegraphics[width=\columnwidth]{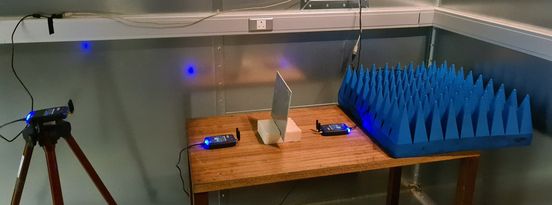}
    \caption{PlutoSDR based NLOS measurement setup in presence of RF absorbers and interferer.}
    \label{fig:meas}
\end{figure}
\begin{figure}
    \centering
    \includegraphics[width=\columnwidth]{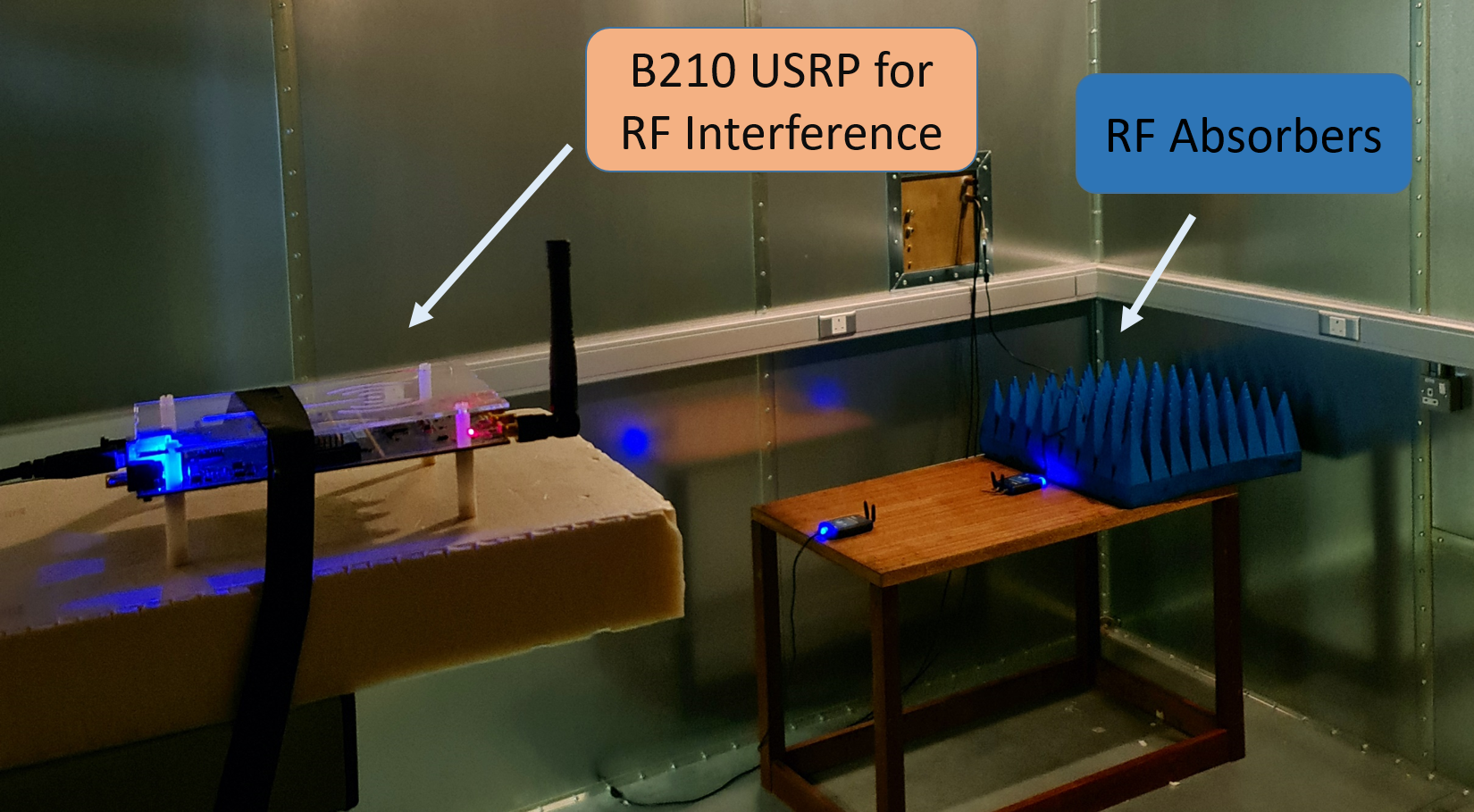}
    \caption{PlutoSDR based LOS measurement setup in presence of wideband RF interference.}
    \label{fig:meas_int}
\end{figure}
Fig. \ref{fig:meas} shows measurement setup where Tx and Rx PlutoSDRs are separated by a NLOS metal sheet. Third PlutoSDR is used as source of narrowband and wideband interference, grid of RF abosrber cones is placed closer to the Tx. We performed measurements in master and scanner configuration. We conducted PER measurement in perfectly metal conducting environment. We observed higher PER values at 2.402 GHz for various Tx and Rx gain values configuration. 
\begin{figure}
    \centering
    \includegraphics[width=\columnwidth]{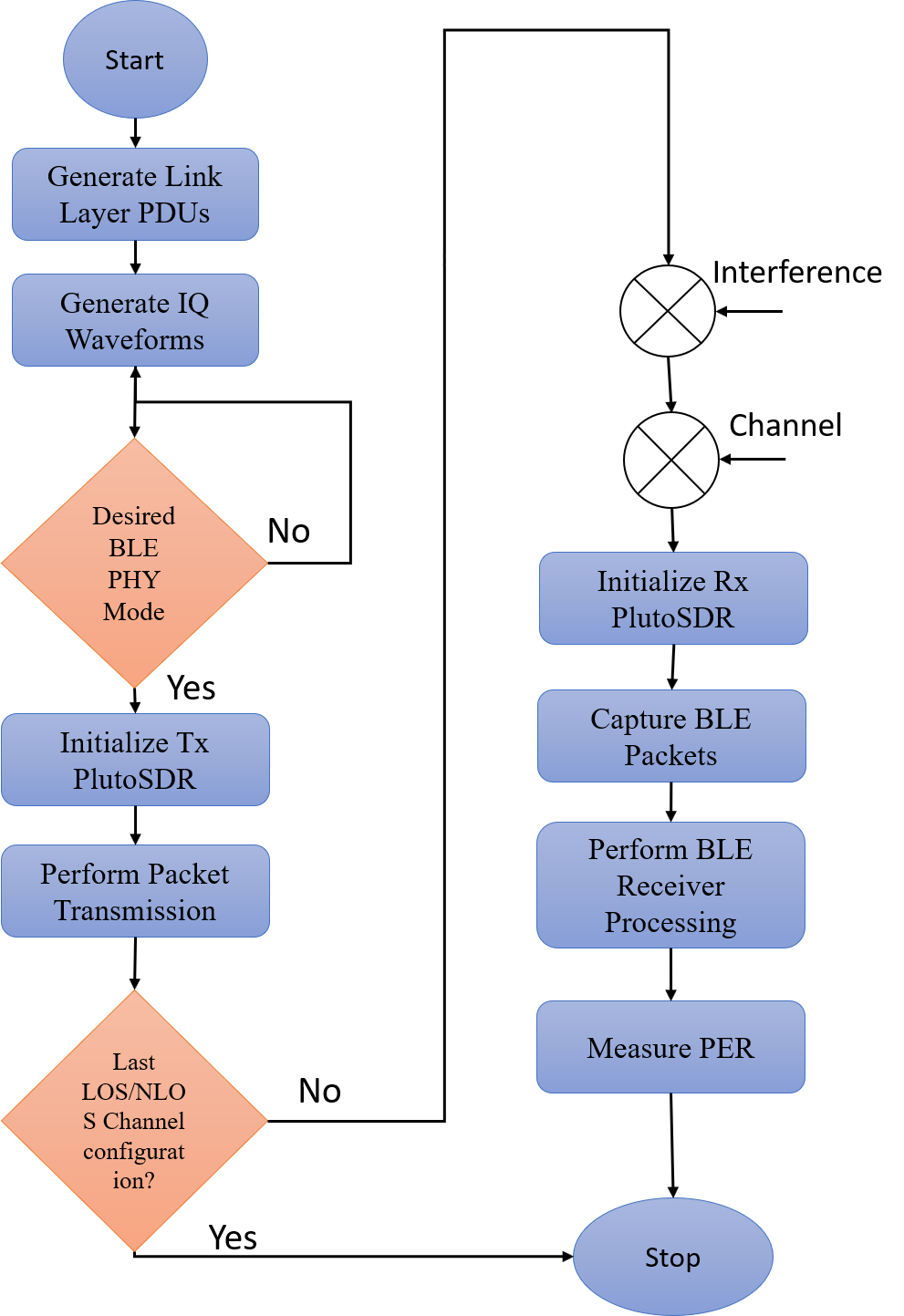}
    \caption{BLE measurement flowchart.}
    \label{fig:flowchart}
\end{figure}
Fig.\ref{fig:flowchart} shows a flowchart of the measurement process and the Tx and Rx processing steps where link-layer PDUs of the desired BLE PHY modes converted into IQ baseband waveform. The baseband IQ waveforms for a given channel configuration are upconverted and transmitted over-the-air using PlutoSDR. BLE waveforms are captured using Rx PlutoSDRs which sends downconverted samples to host PC for receiver processing. After performing receiver operations number of packets and CRC are counted and BLE PER is measured.
\vspace{-0.30cm}
\subsection{LOS and NLOS Scenario}
\begin{figure}
    \centering
    \subfloat[]{\label{fig:tx}\includegraphics[width=0.8\columnwidth]{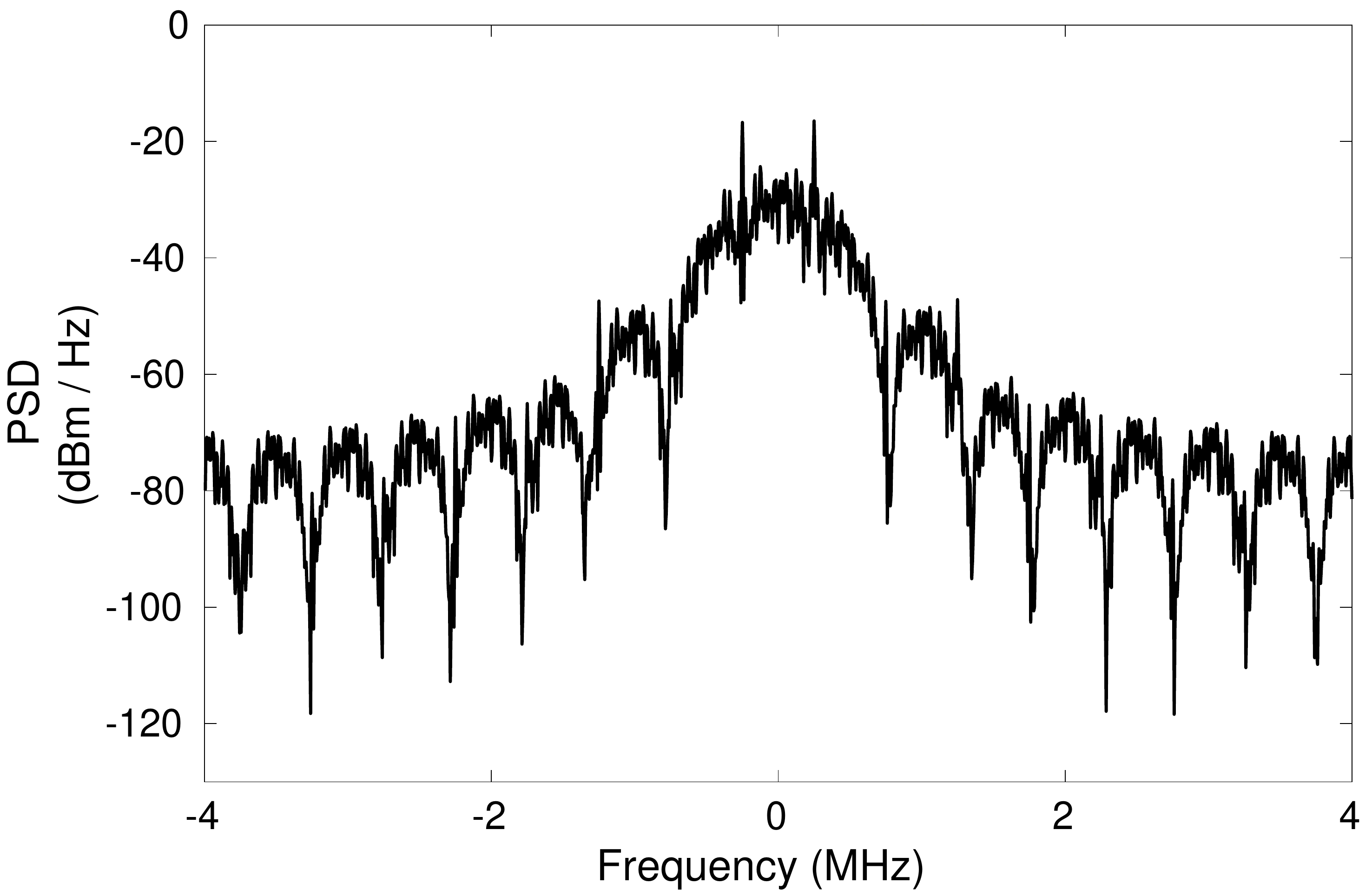}}\\
    \subfloat[]{\label{fig:rx}\includegraphics[width=0.8\columnwidth]{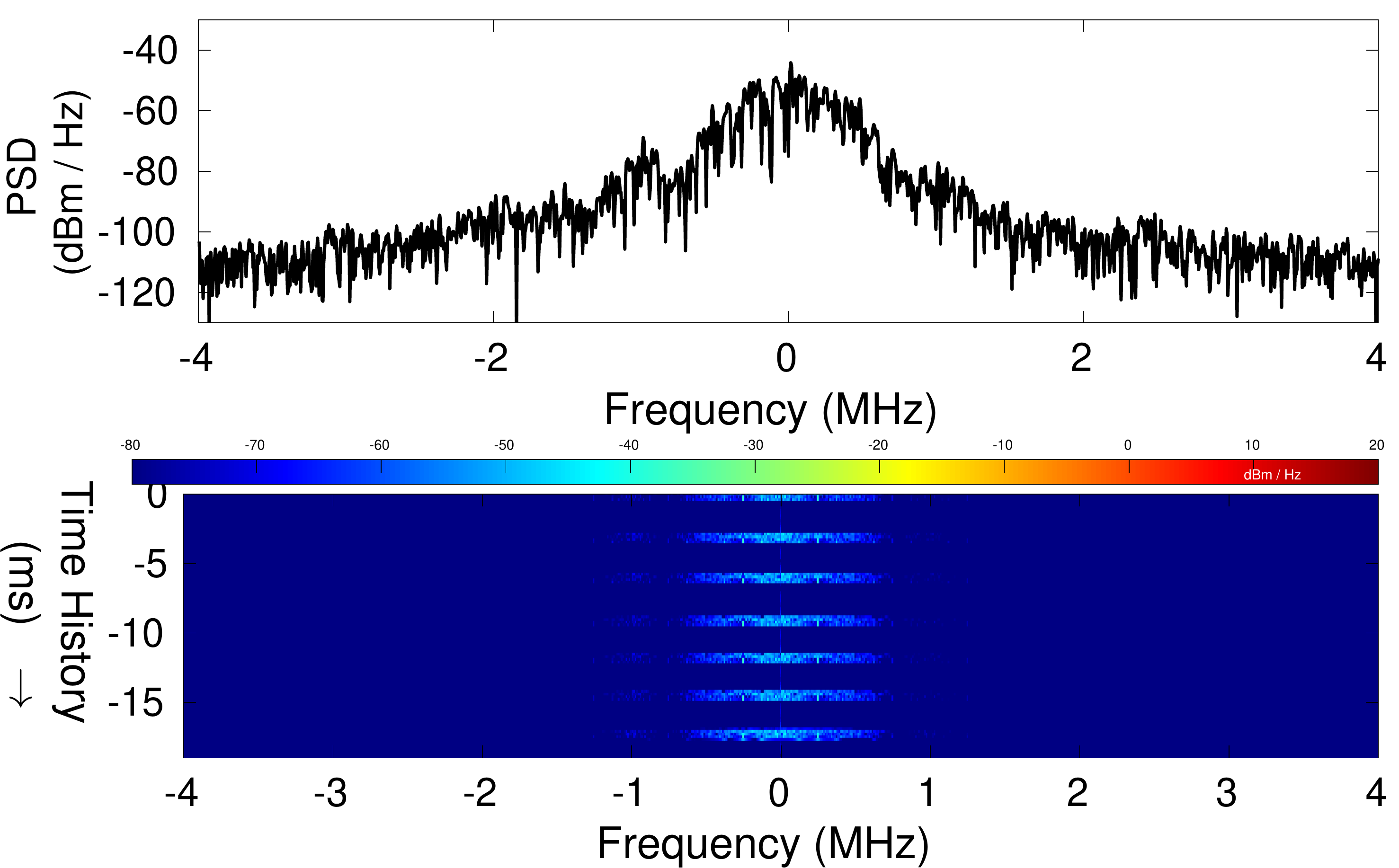}}
    \caption{Baseband spectrum of transmit and received GMSK modulated BLE500K waveform in LOS case for Tx Gain=-10 dB and Rx gain=25 dB.}
    \label{fig:spectrum_spectrogram}
\end{figure}

In this scenario, Tx and Rx PlutoSDRs are placed in LOS and NLOS of each other at a distance of x m. The interferer was inactive in both the LOS and NLOS scenario. We checked the PER measurement at 2.402 GHz, we noticed frequency-selectivity in the baseband spectrum. We conducted PER measurement in a perfectly metal conducting environment. We observed higher PER values at 2.402 GHz for various Tx and Rx gain values configurations. Later on, we introduced RF absorbers to reduce the effect of fading and repeated PER measurements in NLOS for various Tx and Rx gain values. After the PER measurements in NLOS, we repeated the same measurements without the NLOS metal sheet. We observed that no packets were detected, it happened because of frequency selectivity in the spectrum. We changed the Tx and Rx roles of SDR and performed PER measurements in LOS. We noticed that there was no frequency selectivity in the baseband spectrum and the packets were received for all PHY modes of BLE. The addition of absorbers reduced the effect of frequency-selectivity which was observed from the received baseband spectrum. Fig. (\ref{fig:tx} shows clean Tx baseband spectrum of BLE waveform of coded LE125K BLE PHY. At the receiver side, BLE waveform and the packets are captured as shown in Fig.\ref{fig:rx}. 

\begin{figure}
    \centering
    \subfloat[]{\label{fig:nlos}\includegraphics[width=0.8\columnwidth]{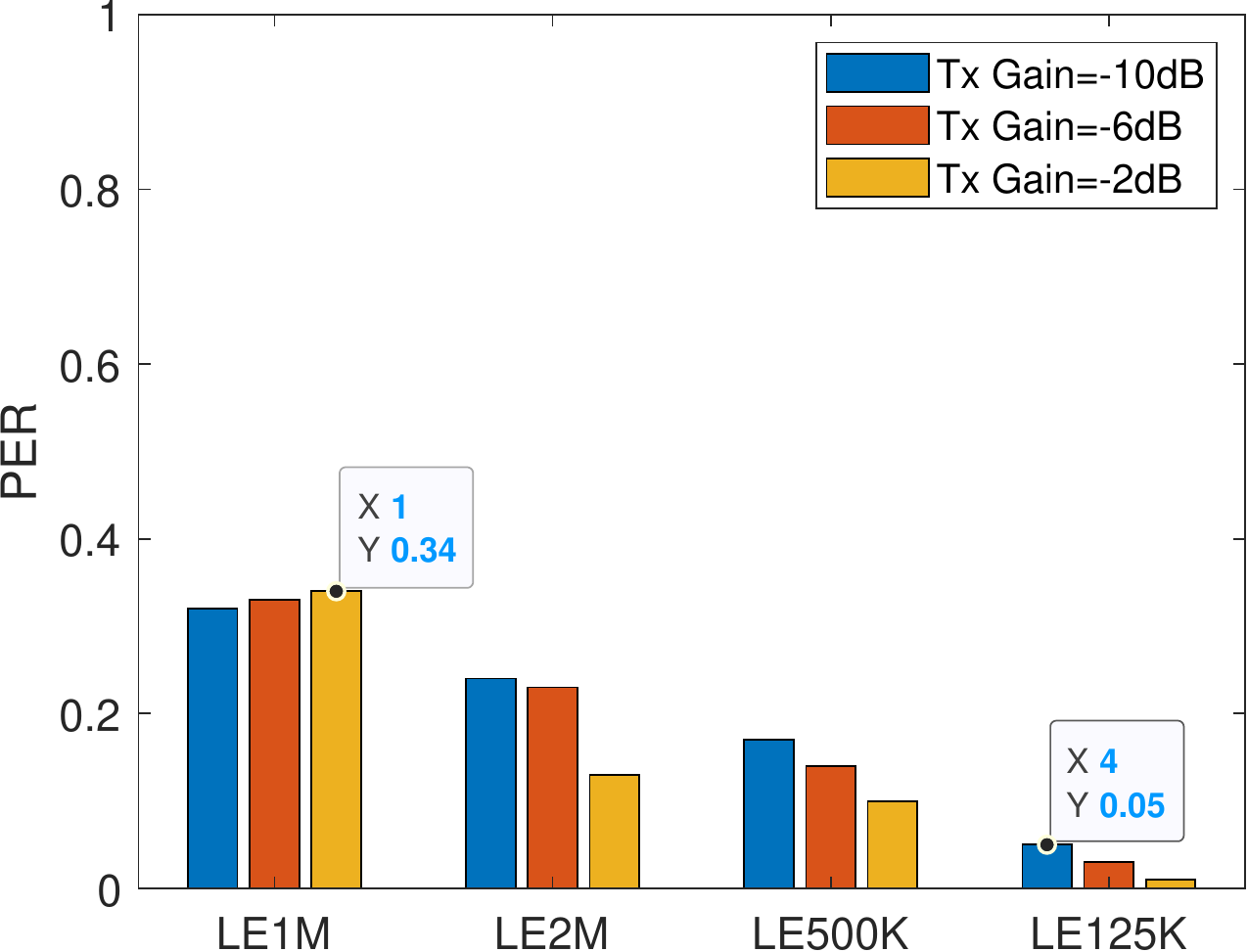}}\\
    \subfloat[]{\label{fig:los}\includegraphics[width=0.8\columnwidth]{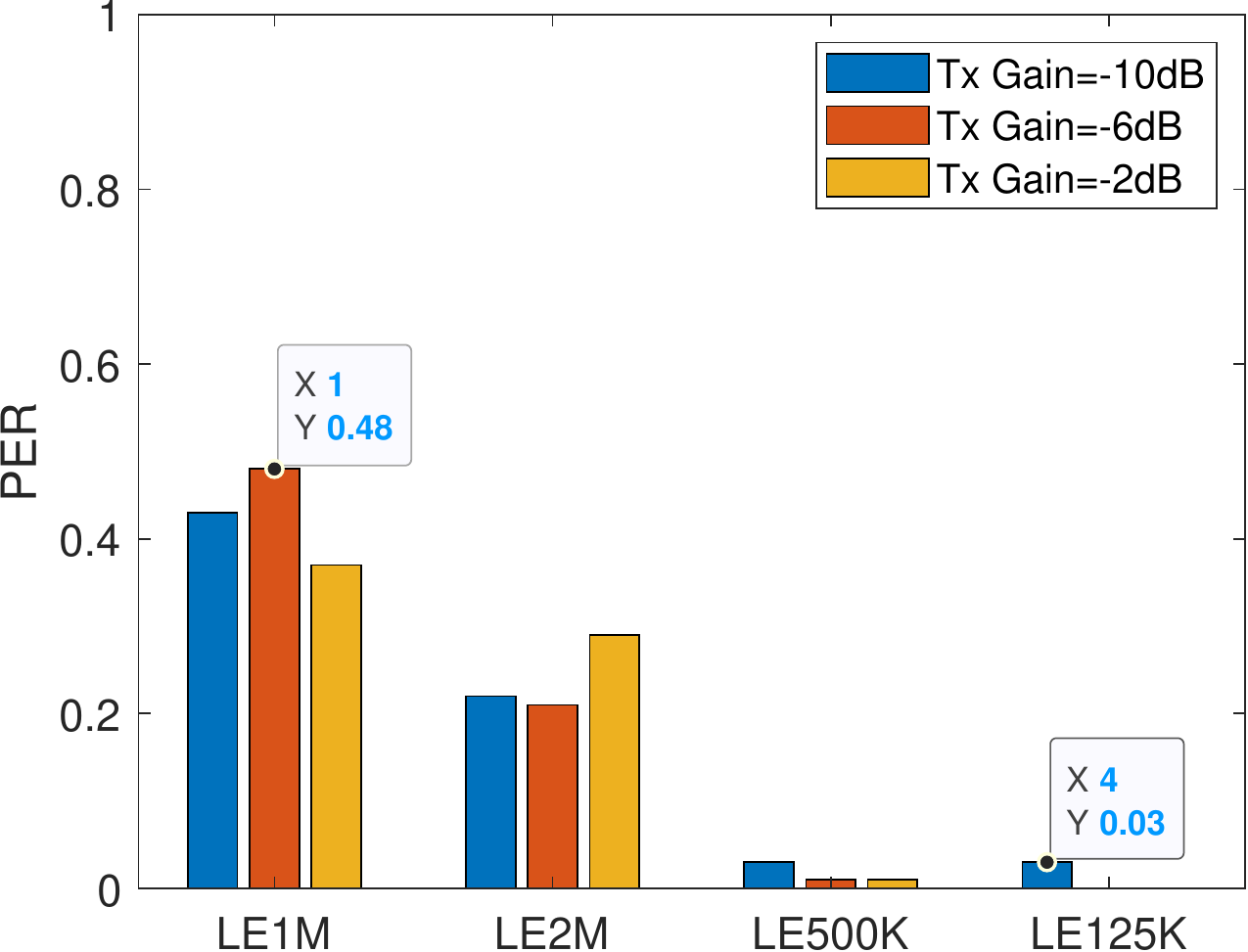}}
    \caption{PER measurement of four BLE PHY in NLOS and LOS channel condition.}
    \label{fig:per}
\end{figure}

 Figure \ref{fig:per} shows PER of BLE PHY modes in NLOS and LOS channel conditions. It can be seen from Fig.\ref{fig:nlos} that the PER is highest for uncoded PHY modes of BLE. A large PER measurement sample was collected for all BLE PHY modes. The PER reduced when we removed the NLOS metal sheet. The reason for reduced PER is the presence of increased forward transmission and dominant LOS component. The existence of a dominant LOS component increased the SNR of the BLE link. In the case of the NLOS metal sheet, there was no forward propagation from Tx to Rx in LOS. PER of coded PHY is lower than the uncoded PHY counterparts in the same channel propagation condition. This is because coded PHY modes are more robust to MPCs. PER for LE125K is the lowest of all BLE PHY modes in both NLOS and LOS scenarios. In the NLOS scenario, the maximum PER has reduced from 0.34 to 0.05 which is an $85.20\%$ reduction in the maximum PER. In contrast to the NLOS scenario, the maximum PER  in the LOS scenario has reduced from 0.48 to 0.03 which is a $93\%$ reduction in the maximum  PER. It is clear from these PER values in both scenarios that the coded PHY modes are robust to fading and offers the lowest PER values in a harsh EM environment.
 \vspace{-0.30cm}
\subsection{LOS and NLOS in Presence of Interference}
\begin{figure}
    \centering
    \subfloat[]{\label{fig:per_65}\includegraphics[width=0.8\columnwidth]{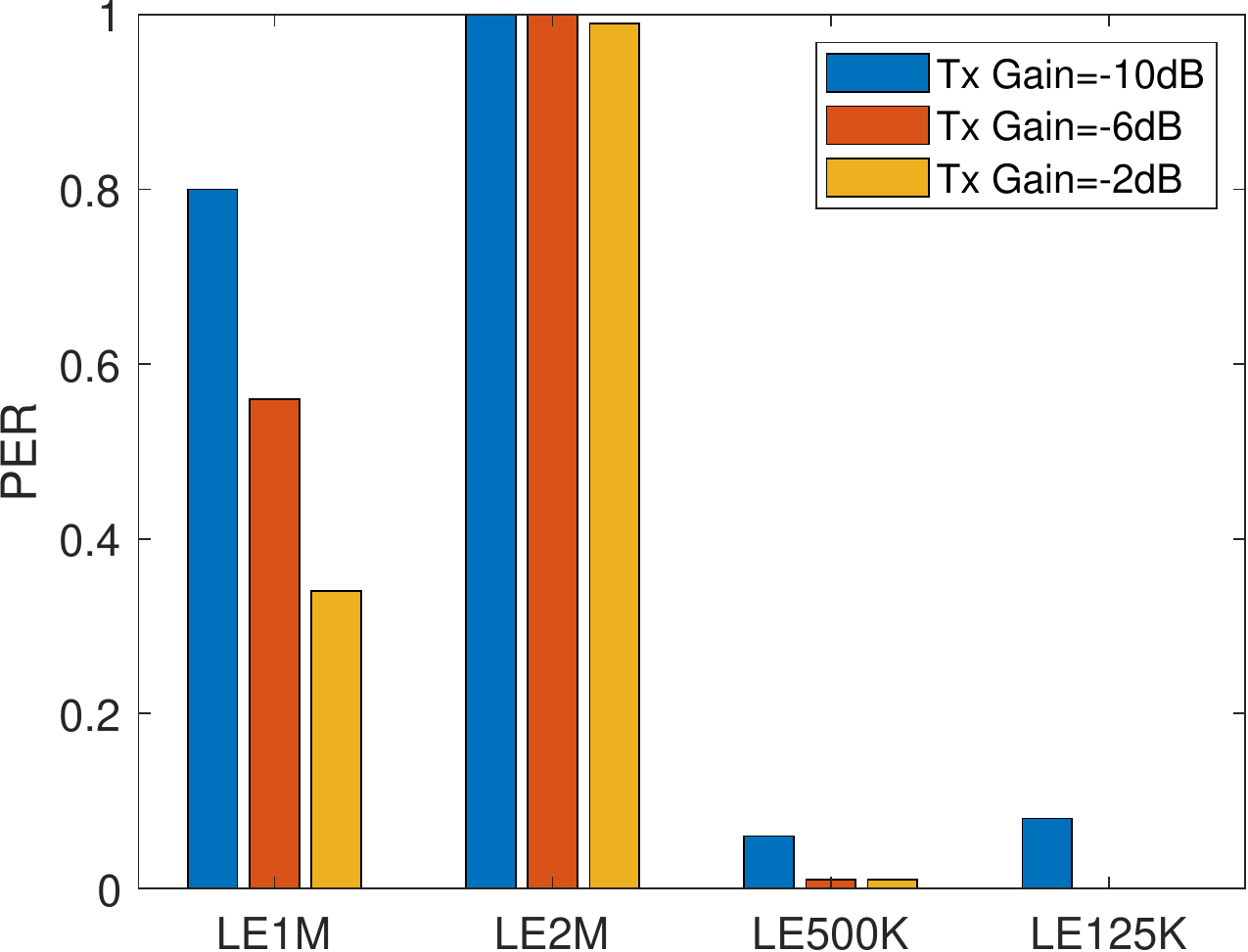}}\\
    \subfloat[]{\label{fig:per_75}\includegraphics[width=0.8\columnwidth]{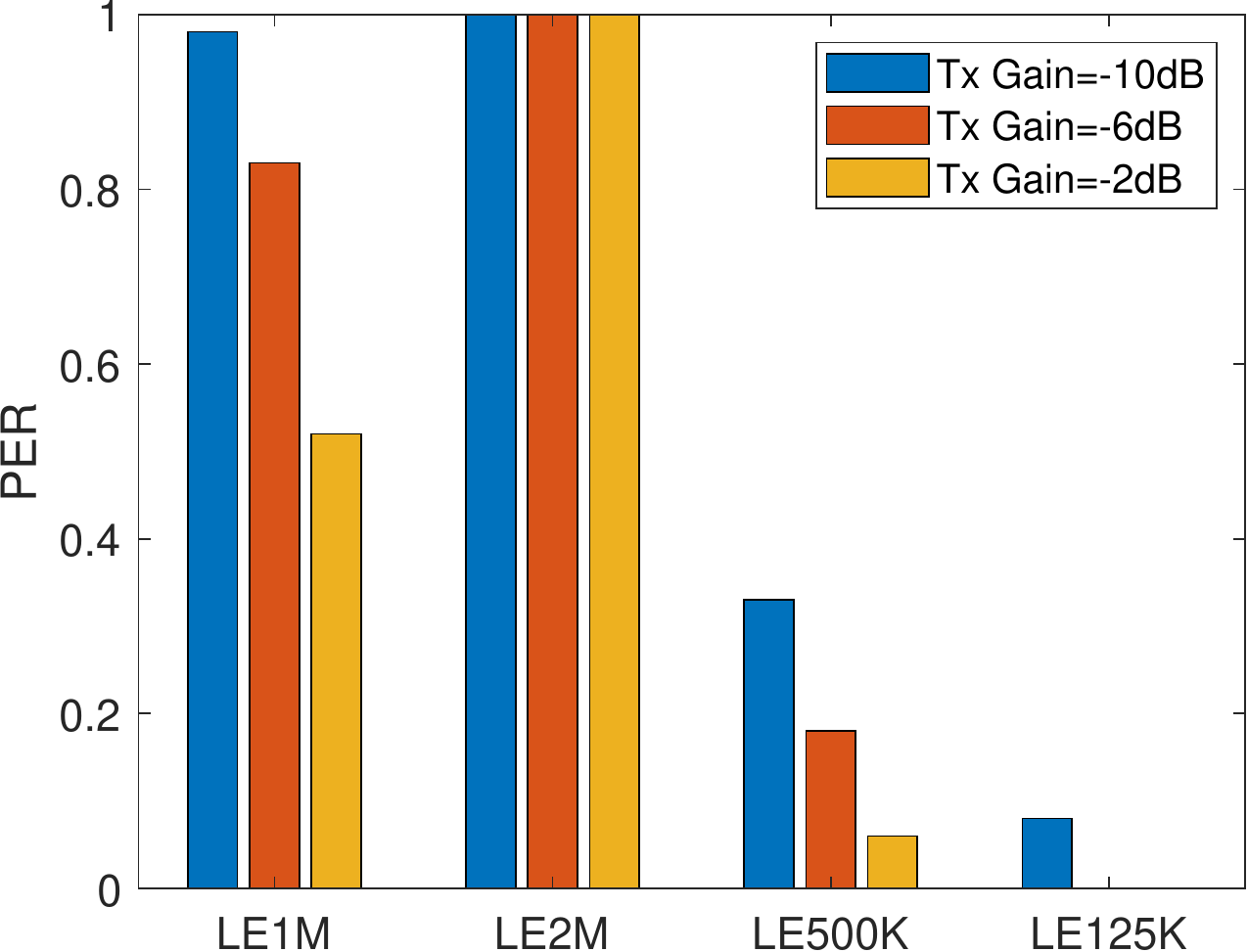}}
    \caption{PER measurement of four BLE PHY in LOS and in presence of WLAN interference.}
    \label{fig:per_with_int}
\end{figure}

This section addresses the PER of BLE in both LOS and NLOS in presence of a WLAN signal operating at a frequency of 2.402 GHz. We generated a 20 MHz wide WLAN interference signal using B210 USRP. B210 USRP was configured to transmit standard-compliant IEEE 802.11a frames \cite{lodro2021near}. We checked the performance in presence of 20 MHz wide WLAN interfering signals. B210 USRP was mounted at around 12 ft from the main BLE communication link. We checked the BLE PER performance for various Tx gain values of B210 USRP. B210 USRP was set to operate with Tx gain values of 65 dB, 70 dB, and 75 dB respectively. For each of the Tx gain values, we checked the PER performance of all BLE PHY modes. We noticed that uncoded BLE PHY modes i.e. LE1M and LE2M exhibited higher PER for all the Tx gain values of the main communication link and at all interference power levels. In contrast to uncoded BLE PHY modes, coded BLE PHY modes LE500K and LE125K were robust to RF interference at all interference power levels.LE1M and LE2M BLE modes were susceptible to interference and the PER was 1 or no packets were detected. In contrast LE1M and LE2M, the LE125K and LE500K were robust to RF interference and have shown low PER in presence of in-band interference (see Fig.\ref{fig:per_with_int}). The worst PER for LE500K was 0.333 and the best was no PERs, where LE125K was highly robust to interference because for most of the runs it has shown no packets received in error.

\vspace{-0.30cm}
\section{Conclusion and Future Work}
We checked the performance of the BLE receiver in the LOS and NLOS scenario in the presence and absence of WLAN interference in a highly reflecting metal environment. Initially, we observed frequency selectivity in the baseband spectrum of BLE received signal both in LOS and NLOS environment. In this configuration, high PER was observed in all BLE PHY modes. Then, we introduced RF absorbers to reduce the effect of frequency selectivity and later performed PER in both LOS and NLOS environments. The addition of RF absorbers removed the frequency selectivity and the PER for BLE PHY modes was improved. After the addition of RF absorbers, we conducted PER measurements in presence of wideband WLAN interference. Higher PER measurements are observed in presence of RF interference for LE1M and LE2M, however, coded LE500K and LE125K are more robust to RF interference.

\vspace{-0.30cm}
\bibliographystyle{IEEEtran}

\bibliography{BLE_PER.bib} 

\begin{thebibliography}{10}
\providecommand{\url}[1]{#1}
\csname url@samestyle\endcsname
\providecommand{\newblock}{\relax}
\providecommand{\bibinfo}[2]{#2}
\providecommand{\BIBentrySTDinterwordspacing}{\spaceskip=0pt\relax}
\providecommand{\BIBentryALTinterwordstretchfactor}{4}
\providecommand{\BIBentryALTinterwordspacing}{\spaceskip=\fontdimen2\font plus
\BIBentryALTinterwordstretchfactor\fontdimen3\font minus
  \fontdimen4\font\relax}
\providecommand{\BIBforeignlanguage}[2]{{%
\expandafter\ifx\csname l@#1\endcsname\relax
\typeout{** WARNING: IEEEtran.bst: No hyphenation pattern has been}%
\typeout{** loaded for the language `#1'. Using the pattern for}%
\typeout{** the default language instead.}%
\else
\language=\csname l@#1\endcsname
\fi
#2}}
\providecommand{\BIBdecl}{\relax}
\BIBdecl

\bibitem{sattler2020risk}
F.~Sattler, J.~Ma, P.~Wagner, D.~Neumann, M.~Wenzel, R.~Sch{\"a}fer, W.~Samek,
  K.-R. M{\"u}ller, and T.~Wiegand, ``Risk estimation of sars-cov-2
  transmission from bluetooth low energy measurements,'' \emph{NPJ digital
  medicine}, vol.~3, no.~1, pp. 1--4, 2020.

\bibitem{cunche2020using}
M.~Cunche, A.~Boutet, C.~Castelluccia, C.~Lauradoux, D.~Le~M{\'e}tayer, and
  V.~Roca, ``On using bluetooth-low-energy for contact tracing,'' Ph.D.
  dissertation, Inria Grenoble Rh{\^o}ne-Alpes; INSA de Lyon, 2020.

\bibitem{leith2020coronavirus}
D.~J. Leith and S.~Farrell, ``Coronavirus contact tracing: Evaluating the
  potential of using bluetooth received signal strength for proximity
  detection,'' \emph{ACM SIGCOMM Computer Communication Review}, vol.~50,
  no.~4, pp. 66--74, 2020.

\bibitem{montoliu2020indoor}
R.~Montoliu, E.~Sansano, A.~Gasc{\'o}, O.~Belmonte, and A.~Caballer, ``Indoor
  positioning for monitoring older adults at home: Wi-fi and ble technologies
  in real scenarios,'' \emph{Electronics}, vol.~9, no.~5, p. 728, 2020.

\bibitem{spachos2020ble}
P.~Spachos and K.~Plataniotis, ``Ble beacons in the smart city: Applications,
  challenges, and research opportunities,'' \emph{IEEE Internet of Things
  Magazine}, vol.~3, no.~1, pp. 14--18, 2020.

\bibitem{ensworth2017ble}
J.~F. Ensworth and M.~S. Reynolds, ``Ble-backscatter: Ultralow-power iot nodes
  compatible with bluetooth 4.0 low energy (ble) smartphones and tablets,''
  \emph{IEEE Transactions on Microwave Theory and Techniques}, vol.~65, no.~9,
  pp. 3360--3368, 2017.

\bibitem{ensworth2017full}
J.~F. Ensworth, A.~T. Hoang, T.~Q. Phu, and M.~S. Reynolds, ``Full-duplex
  bluetooth low energy (ble) compatible backscatter communication system for
  mobile devices,'' in \emph{2017 IEEE Topical Conference on Wireless Sensors
  and Sensor Networks (WiSNet)}.\hskip 1em plus 0.5em minus 0.4em\relax IEEE,
  2017, pp. 45--48.

\bibitem{yu2020plant}
C.~Yu, K.~Kam, Y.~Xu, Z.~Cui, D.~Steingart, M.~Gorlatova, P.~Culligan, and
  I.~Kymissis, ``Plant spike: A low-cost, low-power beacon for smart city soil
  health monitoring,'' \emph{IEEE Internet of Things Journal}, vol.~7, no.~9,
  pp. 9080--9090, 2020.

\bibitem{la2018dense}
Q.~D. La, D.~Nguyen-Nam, M.~V. Ngo, H.~T. Hoang, and T.~Q. Quek, ``Dense
  deployment of ble-based body area networks: A coexistence study,'' \emph{IEEE
  Transactions on Green Communications and Networking}, vol.~2, no.~4, pp.
  972--981, 2018.

\bibitem{luo2019ble}
B.~Luo, F.~Xiang, Z.~Sun, and Y.~Yao, ``Ble neighbor discovery parameter
  configuration for iot applications,'' \emph{IEEE Access}, vol.~7, pp.
  54\,097--54\,105, 2019.

\bibitem{lodro2020near}
M.~Lodro, C.~Smart, G.~Gradoni, A.~Vukovic, D.~Thomas, and S.~Greedy,
  ``Near-field ber and evm measurement at 5.8 ghz in mode-stirred metal
  enclosure.'' \emph{Applied Computational Electromagnetics Society Journal},
  vol.~35, no.~9, 2020.

\bibitem{litwin2001matched}
L.~Litwin, ``Matched filtering and timing recovery in digital receivers,''
  \emph{RF design}, vol.~24, no.~9, pp. 32--49, 2001.

\bibitem{lodro2021near}
M.~Lodro, G.~Gradoni, C.~Smartt, A.~Vukovic, D.~Thomas, and S.~Greedy,
  ``Near-field image transmission and evm measurements in rich scattering
  environment in metal enclosure,'' \emph{Progress In Electromagnetics Research
  M}, vol. 101, pp. 139--147, 2021.

\end{thebibliography}
\end{document}